\documentclass[12pt,a4paper]{elsarticle}
\usepackage{geometry}
\usepackage{setspace}
\usepackage{amsmath}
\usepackage{mathtools}

\usepackage{amssymb}
\begin{document}

\begin{frontmatter}


%and the form \ead[url] for the home page:
% \author{Name\corref{cor1}\fnref{label2}}
% \affiliation{organization={},

\title{Vaporization dynamics of a super-heated water-in-oil droplet: modeling and numerical solution}


\author[inst1]{Muhammad Saeed Saleem}
%\ead{m.s.saleem@utwente.nl}
\affiliation[inst1]{organization={Physics of Fluids Group, Max Planck Center for Complex Fluid Dynamics, University of Twente},%Department and Organization
            addressline={5 Drienerlolaan}, 
            city={Enschede},
            postcode={7522 NB}, 
            country={Netherlands}}

\author[inst1]{Michel Versluis}
\author[inst1]{Guillaume Lajoinie}
\ead{g.p.r.lajoinie@utwente.nl}
\begin{abstract}
The study of vapor bubble growth following droplet vaporization in a superheated liquid involves research areas such as hydrodynamics, heat transfer, mass transfer, and thermodynamics. The interplay between these multiscale aspects is strongly dependent on the geometry, the thermodynamic response, and the local physical properties of the system. To understand the role of each aspect of this complex mechanism we model super-heated droplet vaporization by coupling the equation of motion for bubble growth with the thermodynamics of phase change and heat transfer through the convection-diffusion equation. The semi-analytical model is validated with the analytical description for vapor bubble growth dominated either by inertia (Rayleigh) or by thermal diffusion (Plesset-Zwick), depending on droplet radius and degree of superheat. The effect of a mismatch of the thermal properties between the host liquid and the droplet is shown to be relevant only for low superheating, above which an increase in thermal diffusivity leads to a reduction in the rate of vaporization. At medium to high superheating, the droplet vaporizes completely without relying on thermal diffusion. At the point of complete vaporization, the potential energy within the system drives the bubble overshoots, which vary based on the droplet size and degree of superheat.
\end{abstract}


%%\item Research highlight 1


\end{frontmatter}


\section{Introduction}
\label{sec:sample1}
In the absence of pre-existing gas nuclei, a heated droplet will enter a metastable state where the liquid temperature exceeds its boiling temperature. As temperature further increases, this metastable state ultimately breaks down and results in violent vaporization. Such superheating is commonly observed as water droplets fall into a hot oil-filled pan. There, violent vaporization caused by superheat generates a very characteristic noise and a dreaded oil splash. Industrial devices such as heat exchangers and distillation columns intentionally exploit condensation/vaporization to enhance heat transfer and phase separation. The utilization of droplets dispersed in immiscible hot fluids in a spray column heat exchanger exemplifies direct contact heat transfer, where heat is efficiently exchanged between the dispersed droplets and the hot fluid, enhancing the overall heat transfer efficiency in the system \cite{letan1968mechanism}. Phase separation in a distillation column occurs due to differences in boiling points where water is separated from crude oil by heating the mixture until the water superheats and separates in the form of rising vapor bubbles \cite{mccabe1993unit}.

Two-phase systems where a vapor bubble nucleates and grows in bulk liquid or confined liquid have been well-studied through numerical simulations \cite{can2012level, ory2000growth}, and analytical models with experimental validations \cite{lee1996spherical, dergarabedian1953rate, kosky1968bubble,florschuetz1969growth}. Analytically described by the Rayleigh model \cite{rayleigh1917pressure}, vaporization is initially controlled by inertia. Subsequently, thermal diffusion takes precedence, as described by the Plesset-Zwick \cite{plesset1954growth} model. The recently developed semi-analytical model proposed by Chernov et al. \cite{chernov2020new} aligns well with established classical models, while accurately incorporating both growth dynamics and thermal effects. However, these models do not apply to cases where the infinite bulk medium is replaced with a droplet of finite mass immersed in an infinite immiscible medium. The presence of the host liquid affects the dynamics of the system by modifying inertia, interfacial energy, and viscous dissipation. It also affects heat transfer and limits the amount of liquid available for vaporization. Furthermore, given a spherical geometry, this impact of the host liquid depends on the initial droplet size and on time as the vapor-to-liquid ratio reduces. 

Models have been devolved to approximate heat transfer coefficients. Sideman and Taitel \cite{sideman1964direct} have developed an analytical model that relates the Nusselt number to the Peclet number for droplet radii within the millimeter range. In their system, the droplet vaporizes while rising due to buoyancy through a hot heat transfer fluid. Tochitani et al. \cite{tochitani1977vaporization, tochitani1977vaporizationa} presented a similar relation, but used the Stokes approximation instead of potential flow theory, which extends the validity of the previous model to sub-milimeter droplets. Lajoinie et al. \cite{lajoinie2014ultrafast, lajoinie2020three} have developed three-phase models to describe the vaporization of laser-heated microcapsules. These models, however, use simplified descriptions for heat transfer based on a timescale separation or a simplified thermal boundary layer. This approach may neglect various factors, including temporal and spatial dependencies, and the convective complexities, resulting in less precise representations of the associated heat transfer phenomena. 

Avedisian and Suresh \cite{avedisian1985analysis} developed a semi-analytical model to describe the vaporization process of a droplet in an infinite medium, sustained at a specific superheated temperature. It includes surface tension effects, viscous effects, the inertia of both liquids, and heat transfer. Roesle and Kulacki \cite{roesle2010boiling} extended the model, focusing on dynamics during and after vaporization by employing different governing equations while neglecting the viscous effects. Emery et al. \cite{emery2018bubble} used conservation of mass and energy, neglecting momentum, and this simplified approach overlooks all sources of damping. The main limitation of these three models is the neglecting of acoustic re-radiation as a damping source and equating the gas temperature to boiling temperature throughout the vaporization cycle.%, while the pressure is determined using the ideal gas law. 

In existing literature, models are limited to either estimating heat transfer coefficients or applicable within specific size and temperature ranges. Consequently, they tend to overlook the role of momentum, thereby failing to capture rapid dynamics. In other instances, they are restricted to small sizes, neglecting certain widely recognized sources of damping. Furthermore, a crucial aspect overlooked by all of these studies is the incorporation of convection due to vaporization/condensation during and after vaporization. The dynamics of vapor bubbles at the micro-scale are thus still not fully understood due to numerous divergent and nonlinear effects that do not scale correctly. This is attributed to the intricate interplay of phase change, fluid mechanics, and thermal effects. Modeling and numerically solving vapor bubble growth is essential to shed light on the dominant physical mechanisms driving the vaporization and vapor bubble dynamics as a function of the thermal properties. 

In this work, we aim to develop a model to systematically study the influence of droplet size, degree of superheat, and thermal properties on vaporization dynamics. The proposed semi-analytical model accounts for momentum and associated sources of damping i.e. viscous, surface energy, acoustics, and relates to its vapor temperature that drives the bubble by heat and mass transfer at the interface. We evaluate the gas temperature by accounting for time-dependent convective effects, heat flux, and gas pressure using Antoine’s law. In addition, we account for partially and fully vaporized droplets thus dealing with recondensation. Our numerical results show that the finite vaporizing mass together with the degree of superheat, describes the physical route governing droplet vaporization, i.e. inertia, and thermal diffusion. With a decrease in droplet size and an increase in temperature, the influence of the heat diffusion from the outer medium becomes insignificant. Additionally, both heat diffusion and enthalpy play a crucial role in determining the rate of bubble growth and thus are crucial in controlling bubble expansion. Post-vaporization dynamics are governed by potential energy retained within the system at the time of complete vaporization, and the resulting vapor bubble oscillations are dissipated through acoustic reradiation.

\section{Bubble growth model}
\label{sec:sample2}
This section provides the main steps of the derivation and the details can be found in the appendix. A schematic of the system is shown in Fig. \ref{Sketch}. It consists of a bubble of radius $R_{b}$ nucleating in the center of a water droplet of radius $R_{d}$ placed in oil used as a heat transfer fluid. The entire system is maintained at a homogeneous ambient temperature $T_{\infty}$. Bubble dynamics in an infinite medium is described by the Rayleigh-Plesset equation \cite{rayleigh1917pressure, plesset1949dynamics}. Here, the momentum equation must be integrated across both liquids in a fashion similar to Avedisian and Suresh \cite{avedisian1985analysis} with the addition of an acoustic reradiation term. The resulting Rayleigh Plesset-type equation then takes the following form:

\begin{figure*}[ht]
 \begin{center}
 \includegraphics[width=0.4\linewidth]{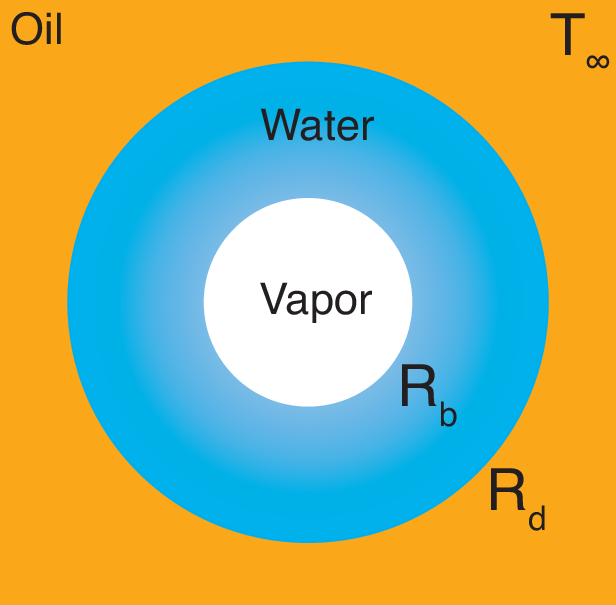} \\ 
 \caption{Configuration of the three-phase system of the present study. A vapor bubble of radius $R_{b}$ nucleates in the center of a water droplet of radius $R_{d}$ placed in oil. The system is maintained at a homogeneous temperature $T_{\infty}$.
 \label{Sketch}}
\end{center}
\end{figure*}

\begin{equation} \label{eq 1}   %%{eq 23}
\begin{split}
(\ddot{R}_{b}R_{b}+2\dot{R}_{b}^2) \left((\rho_{o}-\rho_{w}) \frac{R_{b}}{R_{d}} + \rho_{w}\right) + \frac{(\rho_{w}-\rho_{o}) \dot{R}_{d}^2 }{2} - \frac{\rho_{w} \dot{R}_{b}^2}{2} \\ = P_{g} + \frac{R_{b}}{c_{o}} \dot{P_{g}} - P_{\infty} -2\left(\frac{ \sigma_{wo}}{R_{d}} + \frac{ \sigma_{w}}{R_{b}}\right) -4\left( \frac{\dot{R_{d}}}{R_{d}} (\mu_{o}-\mu_{w}) +\frac{\dot{R_{b}}}{R_{b}} \mu_{w} \right), 
\end{split}
\end{equation}
where $r$ is the radius, $\rho$ is the density, $\sigma$ is the interfacial tension, $\mu$ is the viscosity, $P$ is the pressure, and $c$ is the speed of sound. The subscripts $b, d, w, o, g,$ and $\infty$ represent the bubble, droplet, water, oil, gas, and ambient conditions far away from the bubble (at infinity), respectively. The single and double overdots indicate the first and second time derivatives, respectively. The bubble is driven by a pressure difference between the gas core and the surrounding liquid. The vapor saturation pressure is computed using the semi-empirical Antoine law \cite{antoine1888nouvelle} that describes the liquid-vapor transition:

\begin{equation}\label{eq 2}    %%{eq 24}
P_{g}  = 10^ {5+A - \frac{B}{C+T_{g}}},
\end{equation}
where, $A, B,$ and $C$ are the Antoine coefficients and $T_{g}$ represents the temperature of the gas. The pressure in the bubble depends on its temperature, vapor mass, and bubble size. The bubble size is ultimately governed by the momentum conservation equation Eq. \ref{eq 1}, while the mass changes through vaporization/condensation and depends on convection, specific heat of the vaporizing liquid, on the vaporization enthalpy, heat conduction across the interface, and mass available for vaporization. Taking the time derivative of the perfect gas law, eliminating the gas pressure term through Antoine's law, and considering the heat flux across the interface yields the equation that governs the evolution of the gas temperature:

\begin{equation}\label{eq 3}    %{eq 32}
\dot{T_{g}} \Bigg[\frac{1}{T_{g}} - \frac{Bln(10)}{(C+T_{g})^2} - \frac{C_{p_{g}} }{H_{w}}\Big(1+\frac{\rho_{g}}{\rho_{w}}\Big) \Bigg] =  \frac{3 \dot{R_{b}}}{R_{b}} - \frac{{4 \pi R_{b}^{2}k_{w}}}{H_{w}m_{g}}  \frac{\partial T}{\partial r} \bigg|_{R_{b}} \Big(1+\frac{\rho_{g}}{\rho_{w}}\Big),
\end{equation}
where, $C_{p}$ is the specific heat, $H$ is the enthalpy of vaporization, $k$ is the thermal conductivity, $m$ is the vaporized mass. As the bubble expands it also consumes mass at the interface through vaporization which can be expressed by writing the heat transfer across the interface:  

\begin{equation}\label{eq 4}    %{eq 28a}
\dot{m_{g}} = \frac{{4 \pi R_{b}^{2}k_{w}}}{H_{w}}  \frac{\partial T}{\partial r} \bigg|_{R_{b}} - \frac{m_{g} C_{p_{g}} \dot{T_{g}}}{H_{w}},   
\end{equation}
where, $r$ is the spatial coordinate, $T$ is temperature outside of bubble. When the droplet is fully vaporized, no more liquid can undergo phase change and the vapor and liquid phases no longer coexist. The vapor is then assumed to behave as an ideal gas such that:

\begin{equation}\label{eq 5}   %{eq 30}
\dot{P_{g}} = P_{g} \Big(\frac{ \dot{T_{g}}}{T_{g}} - \frac{3 \dot{R_{b}}}{R_{b}}\Big).
\end{equation}

The choice of a differential form avoids the creation of an artificial discontinuity in the vapor behavior as the system transitions from a partially vaporized core to a fully vaporized one. Heat transferred to the bubble is now used only to adjust the gas temperature and its pressure through the bubble volume (see Eq. \ref{eq 5}). The gas temperature is then given by:

\begin{equation}\label{eq 6}    %{eq 33}
\dot{T_{g}}= \frac{{4 \pi R_{b}^{2}k_{w}}}{C_{p_{g}}m_{g}}  \frac{\partial T}{\partial r} \bigg|_{R_{b}},
\end{equation}
since $\dot{m_{g}}= 0$. The temperature outside the bubble in both cases is determined by solving the convection-diffusion equation:

\begin{equation}\label{eq 7}   %{eq 34}
\dot{T}(r,t) = \frac{D}{R} \frac{\partial^2 (RT(r,t))}{\partial r^2}  - \dot{R} \frac{\partial T (r,t)}{\partial r},
\end{equation}
where $R$ corresponds to a bubble or drop. In their respective domains (water or oil) $D$ denotes the thermal diffusivity.

\section{Numerical integration}
\label{sec:sample3}

\begin{table}[!h]
\begin{center}
\begin{tabular}{|c c|} 
 \hline
Parameters & Initial conditions \\ [0.5ex] 
 \hline\hline
  Temperature & $T(r,0) = T_{\infty}$ \\
\hline
Pressure &  $P_{g}(0)  = 10^ {5+A - \frac{B}{C+T_{\infty}}},$ \\
\hline
Initial bubble &  $R_{b}(0) = 2.5 \mu m$ \\
\hline
Bubble velocity &  $\dot{R_{b}}(0) = 0 $ \\
\hline
Bubble mass & $m_{g}(0) = \frac{4}{3} \pi R_{b}^3 (0) \frac{P_{g}(0) M_{n}}{R_{i}T_{\infty}} $ \\
\hline
Droplet radius &  $R_{d}(0) = (R_{d}(0)^3 + R_{b}(0)^3)^\frac{1}{3} $ \\
\hline
Droplet mass &  $m_{w}(0) = \frac{4}{3} \pi R_{d}^3 \rho_{w} $ \\
\hline
 \hline\hline
Parameters & Boundary conditions \\ [0.5ex] 
 \hline\hline
Droplet Radius &  $R_{d} (t) = \left( \frac{3 (m_{w}-m{g})}{4 \pi \rho_{w}} + R_{b}^3 \right) ^{\frac{1}{3}}$ \\
 \hline
 Droplet velocity &  $\dot{R_{d}}(t) = \frac{\dot{m_{g}}}{4\pi\rho_{w}R_{d}^{2} } + \dot{R_{b}} \bigl(\frac{R_{b}}{R_{d}}\bigl)^{2}$ \\
 \hline
Interfaces: &  \\
Vapour-Water & $T(r = R_{b},t) = T_{g}$ \\
Water-Oil &  $k_{w} \frac{\partial T}{\partial r} \big|_{_{w}} = k_{o} \frac{\partial T}{\partial r} \big|_{_{o}} $ \\
Ambient & $T(r = r_{\infty},t) = T_{\infty}$ \\
\hline

\end{tabular}
\caption{\label{table 1}Initial and boundary conditions.}
\end{center}
\end{table}

In the vaporization regime, the momentum is governed by Eq. \ref{eq 1}, and the vapor pressure is determined by Antoine's law Eq. \ref{eq 2} which requires knowing the vapor temperature. This temperature is computed using equation \ref{eq 3}, and mass transfer using Eq. \ref{eq 4}. The temperature profile outside of the bubble is determined by Eq. \ref{eq 7}. After vaporization, Eq. \ref{eq 5} is used to calculate the vapor pressure, and the vapor temperature is computed using Eq. \ref{eq 6}.
\begin{table}[!h]
\begin{center}
\begin{tabular}{|c c c|} 
 \hline
 Parameter & Value & Unit \\ [0.5ex] 
 \hline\hline
 Antoine coefficient A$^+$ & 5.08354 & - \\ 
 \hline
 Antoine coefficient B$^+$ & 1663.125 & - \\ 
 \hline
 Antoine coefficient C$^+$ & -45.622 & - \\ 
 \hline
 Dynamic viscosity $(\mu_{w})^+$ & function of temp. \cite{nistlink} & $Pa.s$ \\
 \hline
 Dynamic viscosity $(\mu_{o})^*$ & function of temp. \cite{nistlink} & $Pa.s$ \\
 \hline
 Thermal conductivity $k_{w}^+$ & 0.61 & $W.m^{-1}.K^{-1}$ \\
  \hline
 Thermal conductivity $(k_{o})^*$ & 0.13 & $W.m^{-1}.K^{-1}$ \\
 \hline
 Density $(\rho_{w}) ^+$ & 998 & $kg.m^{-3}$ \\
 \hline
 Density $(\rho_{o}) ^*$ & 789 & $kg.m^{-3}$ \\
  \hline
 Specific heat $(C_{p_{w}})^+$ & 4216 & $J.kg^{-1}.K^{-1}$ \\
 \hline
  Specific heat $(C_{p_{g}}) ^\times$ & 2000 & $J.kg^{-1}.K^{-1}$ \\
 \hline
 Specific heat $(C_{p_{o}})^*$  & 2272 & $J.kg^{-1}.K^{-1}$ \\
  \hline
 Vaporization enthalpy $(H_{w}) ^+$ & function of temp. \cite{nistlink} & $J.kg^{-1}$ \\
 \hline
Surface tension $(\sigma_{w})^{+\times}$ & function of temp. \cite{nistlink} & $N.m^{-1}$ \\
 \hline
Surface tension $(\sigma_{o})^{*}$ & function of temp. \cite{STocta} & $N.m^{-1}$ \\
 \hline
Interfacial tension $(\sigma_{wo})^{+*}$ & $\sigma_{w} - \sigma_{o}$ & $N.m^{-1}$ \\
 \hline
 Atmospheric Pressure $(P_{\infty})$ & 101325 & $Pa$ \\
  \hline
 Speed of sound $(c_{o})^*$ & 1067 & $m.s^{-1}$ \\
  \hline
  Molecular mass $(M_{n})^+$ & 18 x 10 $^{-3}$ & $kg.mol^{-1}$ \\
  \hline
Ideal gas constant $(R_{i})$ & 8.314 & $J.K^{-1}.mol^{-1}$ \\
  \hline
Saturation temperature $(T_{s})$ & 373 & $K$ \\
  \hline
\end{tabular}
\caption{\label{table 2}Material properties: $^+$ Water-liquid, $^\times$ Water-vapor, $^*$ Octadecene}
\end{center}
\end{table}
This set of coupled differential equations is integrated in MATLAB using the ode 45 solver to compute the bubble radius. The convection-diffusion equation (Eq. \ref{eq 7}) is solved by the method of line where the time dependency is handled by the solver while the spatial dependency is discretized using a re-centered finite difference scheme on the Eulerian grid. The grid vector $r$ is defined as:

\begin{equation}\label{eq 8}
r =  \left[ 0, \sum_{i=1}^{N} d_{i} \right],
\end{equation}
where $d = (1,2,...,N)^{(p-1)}r_{1} $, $N = (\frac{r_{m}p}{r_{1}}) ^{1/p}$, $r_{1} = 1.9 \times 10^{-21}$ $m$, $r_{m} = 0.1$ $m$, $p = 4$. Note that, if the pressure drops below the saturation point after full vaporization, the system can return to a partially vaporized state and resume vaporization or condensation. The initial conditions and boundary conditions are listed in Table \ref{table 1} and values of all parameters are listed in Table \ref{table 2}. We have used octadecene (oil phase) as the heating medium. The model is solved for a range of droplet sizes ($R_{d}$) ranging from $5$ to $1000$ $\mu m$ and degrees of superheat $(T_{\infty} - T_{s})$ ranging from $15$ to $250$ $^{\circ}C$.

\section{Results}
\label{sec:sample4}

\subsection{Model validation}   \label{Model validation}
To validate our model, we have compared bubble growth to analytical benchmarks describing the growth of a vapor bubble in an infinite pool of its own liquid in a free field. The system is maintained at an ambient temperature $(T_{\infty})$ and atmospheric pressure $(P_{\infty})$. Upon nucleation, the bubble starts to grow in an inertia-dominated regime described by the Rayleigh \cite{rayleigh1917pressure} model with $R_{b} \sim t$. It then transitions towards a thermal diffusion-dominated growth described by the Plesset-Zwick \cite{plesset1954growth} model $R_{b} \sim t^{1/2}$. Mikic et al. \cite{mikic1970bubble} combined these two models to capture both regimes in their respective time domains:

\begin{equation}\label{eq 9}   %{eq 34}
R^+ = \frac{2}{3} \left[ (t^+ + 1)^{3/2} - (t^+)^{3/2} - 1 \right],
\end{equation}
with 
\begin{equation}\label{eq 10}   %{eq 34}
t^+ = \frac{\pi}{18} \frac{H_{w} (T_{\infty} - T {s}(P_{\infty}))\rho_{g}}{T_{s}(P_{\infty}) \rho_{w}} \frac{t}{Ja^{2} D_{w}},
R^+ = \frac{\pi}{12} \sqrt{\frac{2H_{w} (T_{\infty} - T {s}(P_{\infty}))\rho_{g}}{3 T_{s} \rho_{w}}} \frac{R_{b}}{Ja^{2} D_{w}},
\end{equation}
\begin{figure*}[ht]
 \begin{center}
 \includegraphics[width=1\linewidth]{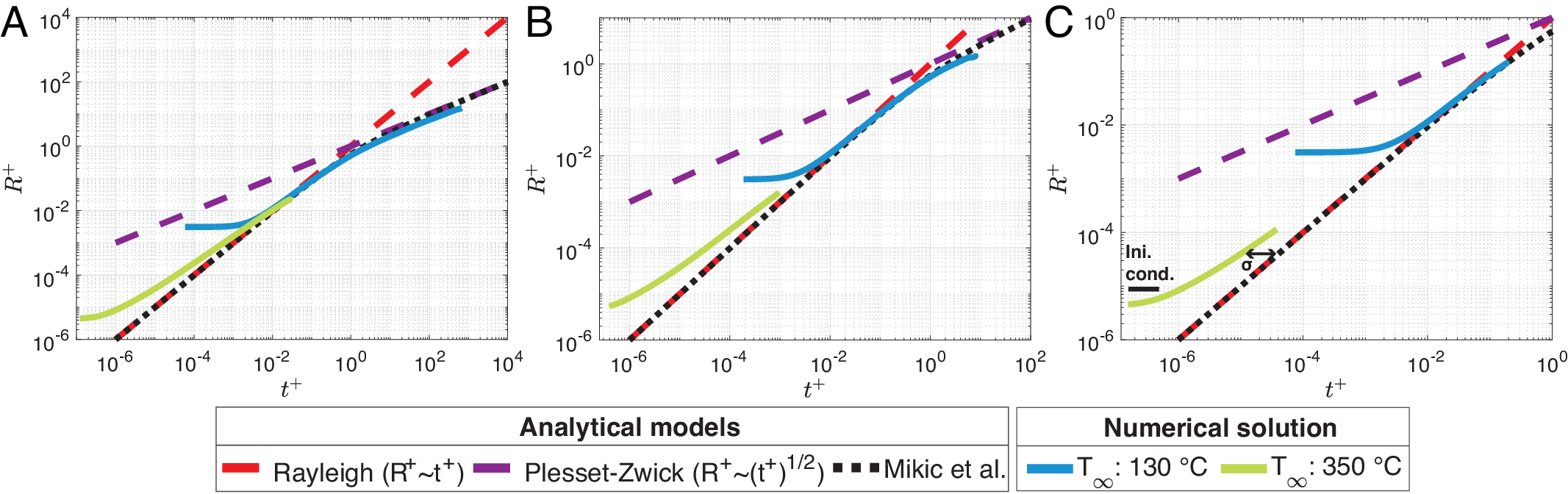} \\ 
 \caption{Verification of the numerical results and comparison of the proposed semi-analytical model to classic analytical models: A. $R_{d}:$ $1000$ $\mu m$. B. $R_{d}:$ $100$ $\mu m$. C. $R_{d}:$ $10$ $\mu m$.     
 \label{analytical_comp}}
\end{center}
\end{figure*}
where $Ja = \frac{\rho_{w} c_{p_{g}} (T_{\infty} - T {s}(P_{\infty}))}{\rho_{g} H_{w}}$ is the Jakob number and the subscript $s$ represents saturation conditions. To compare our model with this classical two-phase configuration: we first consider a water droplet of radius $1$ $mm$, (i.e., infinite with respect to the inertia and diffusive boundary layers over the simulated time-span). The drop is still theoretically immersed in oil, but its size also renders surface tension effects negligible. 

The analytical result of Eq. \ref{eq 9} can be seen in Fig. \ref{analytical_comp} with a comparison to our model’s prediction (non-dimensionalized by Eq. \ref{eq 10}) for ambient temperatures of $130$ and $350$ $^{\circ}C$. Our model is in excellent agreement with the analytical solutions, except for a short initial time characterized by a slow growth in the $R^{+}$ versus $t^{+}$ curves. This regime corresponds to a surface-tension-dominated growth that is not considered in the analytical models.
It is also clear from Fig. \ref{analytical_comp} that the degree of superheat has a strong influence on the vaporization regime following the initial dynamics dominated by surface tension: increased superheat strongly favors inertial growth.

In Fig. \ref{analytical_comp}B and Fig. \ref{analytical_comp}C, Eq. \ref{eq 9} and our model (plotted through Eq. \ref{eq 10}) are evaluated for bubble sizes of $100 \mu m$, and $10 \mu m$, respectively, and ambient temperature of $130$ $^{\circ}C$ and $350$ $^{\circ}C$. The main features remain similar to the case of the 1-$mm$ droplet, and the bubble growth still largely follows the analytical prediction even if it is evaluated beyond its area of validity i.e. for medium and small droplets. There are, however, several noteworthy consequences. First, the smaller the droplet, the more dominant inertia is: both droplets indeed vaporize almost entirely within the inertial regime. Second, the degree of superheat has a stronger influence on smaller droplets, as evidenced by the increasing separation between the blue and green curves. Finally, surface tension (not included in the analytical models) becomes important for smaller droplets, which results in a significant mismatch between the result of Eq. \ref{eq 9} and our model for high degrees of superheat.

\subsection{Initial bubble growth}    \label{Growth dynamics of vapor bubble}
\begin{figure*}[ht]
 \begin{center}
 \includegraphics[width=1\linewidth]{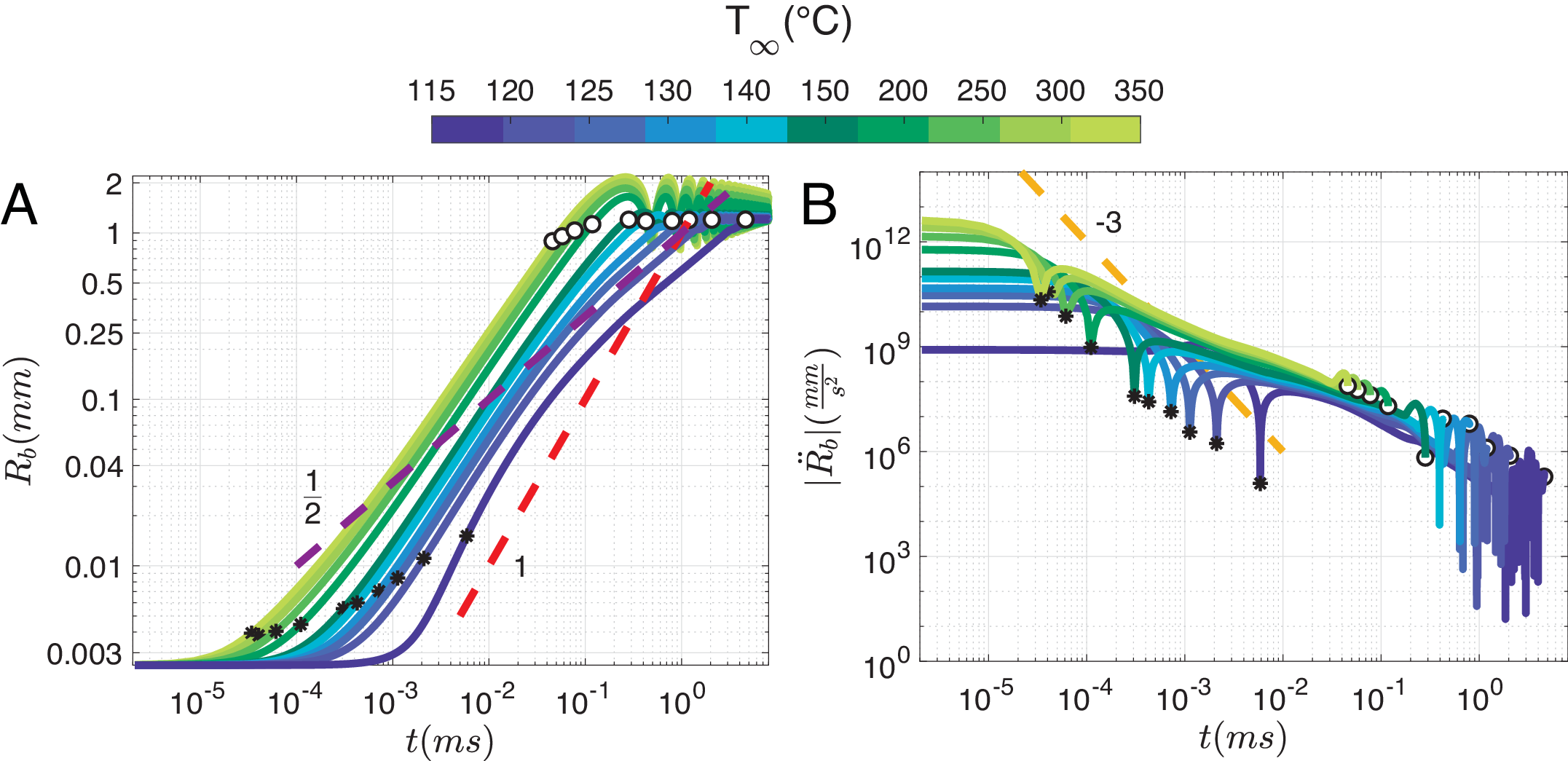} \\ 
 \caption{Vaporization dynamics of a $100$ $\mu m$ radius droplet at different ambient temperatures. Evolution of: A. bubble radius, B. bubble wall acceleration. Circles and asterisks indicate full vaporization and maximum velocity respectively.
 \label{model-results}}
\end{center}
\end{figure*}

The temporal evolution of the bubble radius resulting from the vaporization of the $100$ $\mu m$ droplet for a range of ambient temperatures is shown in Fig. \ref{model-results}A. Initially, the growth of the nucleated bubble remains stagnant, after which it grows homogeneously inside the droplet until complete vaporization, denoted by the circle on the radius-time plot. Post-vaporization, the bubble undergoes damped oscillations with an amplitude that diminishes with a decrease in ambient temperature, as discussed in section \ref{post-vap}. The initial growth of the bubble is governed by gas pressure and inertia:
%In the preceding section, we observed how complex bubble growth dynamics can be simplified into power-law-based physical descriptions based on classical analytical models. Similarly, we will now describe the initial characteristics of bubble growth through the Rayleigh model without the influence of any damping source: 

\begin{equation}\label{eq 11}  
\ddot{R}_{b} = \frac{\frac{\Delta P }{\rho_{l}}R_{b_{0}}^{2} - (\dot{R}_{b_{0}}R_{b_{0}})^{2}}{R_{b_{0}}^{3}},
\end{equation}
where $\Delta P$ is the difference in pressure between the bubble and liquid, and the subscript $0$ denotes the initial conditions. The initial velocity of the bubble is kept at zero to let the bubble expand using energy stored within the system. With $\dot{R}_{b_{0}} = 0$, Eq. \ref{eq 11} can be simplified to $\ddot R_{b} \sim \frac{\Delta P}{\rho_{l}} \frac{1}{R_{b_{0}}}$. The bubble is thus driven by pressure difference and maintains a constant acceleration as long as $R_{b}$ = $R_{b_{0}}$. In Fig. \ref{model-results}B, the plot shows the evolution of the bubble wall acceleration, corresponding to the period during which initial conditions dominate bubble growth. The driving pressure can be estimated through Taylor expansion as:

\begin{equation} \label{eq 13} 
\Delta P = \begin{cases*}
   P_{sat}(T_{\infty}) - P_{\infty}, & $>$ 10$^{o}C$ superheat,\\
  (T_{g} - T_{\infty}) \frac{\partial P}{\partial T_{g}} (T_{\infty}), & near boiling.
\end{cases*}
\end{equation}

In our simulation, we consider superheating typically exceeding 10$^{o}C$, which drives the system through the difference between saturation pressure and ambient pressure. Integrating and substituting zero velocity in the Rayleigh-Plesset model, the initial growth follows $R_{b_{0}}^{3}= \bigl({\frac{\Delta P}{\rho_{l}} t^{2} }\bigr)^\frac{3}{2}$, substituted in Eq. \ref{eq 11}:

\begin{equation}\label{eq 12}  
\ddot{R}_{b} = \Bigl(\frac{\rho_{l}}{\Delta P} \Bigr) ^{\frac{1}{2}} R_{b_{0}}^{2} t^{-3}.
\end{equation}

A short transition from the effect of initial conditions towards natural bubble growth occurs approximately around $t = R_{b_{0}} \sqrt{\frac{\rho_{l}}{\Delta P}}$, which would decrease for a higher degree of superheat. Afterward, the bubble follows either $(R \sim t)$ or $(R \sim t^{1/2})$ described by inertial or thermal growth as discussed in section \ref{Model validation} and shown by the dashed line in Fig. \ref{model-results}A.

\subsection{Dominant phenomena}
\begin{figure*}[b!]
 \begin{center}
 \includegraphics[width=1\linewidth]{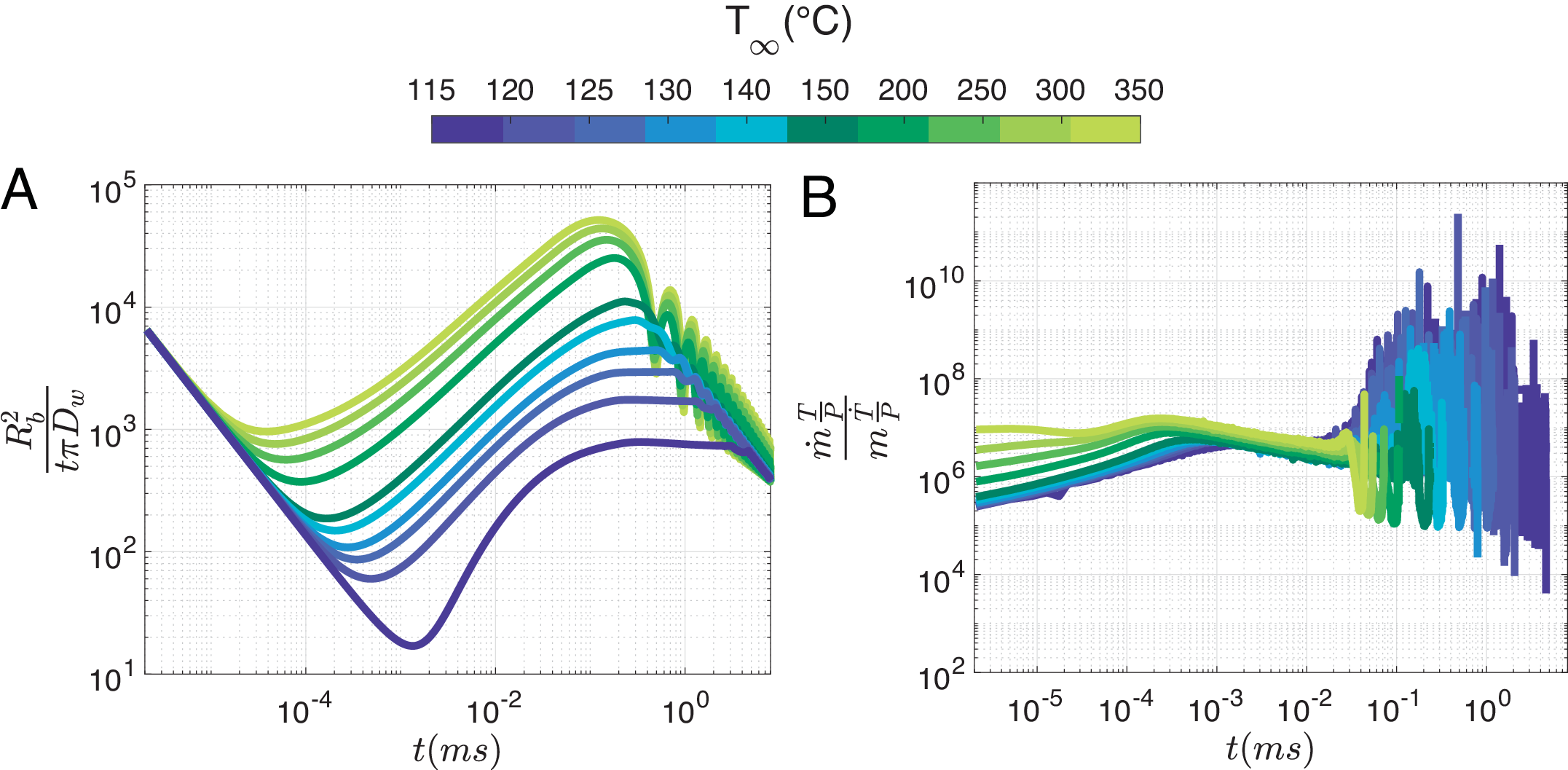} \\ 
 \caption{Vaporization of a $100$ $\mu m$ radius droplet. Evolution of: A. unsteady state of bubble growth, B. ratio of the mass transfer versus gas temperature. 
 \label{model-resultsa}}
\end{center}
\end{figure*}

During vaporization heat diffuses from the surroundings towards the bubble through its thermal boundary layer, which typically follows \cite{lajoinie2014ultrafast}: 

\begin{equation}\label{eq 14}  
\frac{1}{\delta_{th}} = \frac{1}{\sqrt{\pi D_{w} t}} + \frac{1}{R_{b}}.
\end{equation}
where $\delta_{th}$ is the thickness of the thermal boundary layer, and $\tau_{th} = \frac{R_{b}^2}{\pi D_{w}}$ denotes the corresponding time scale for the development of the boundary layer. The plot in Fig. \ref{model-resultsa}A depicts the time scale $\frac{\tau_{th}}{t}$, where the ratio significantly exceeds one. This suggests that the entire phase change event occurs out of equilibrium, primarily due to rapid vaporization. As heat diffuses it is consumed in phase change and gas heating. The increase in the volume of gas is therefore related to both the rate of mass transfer and the rate of change in gas temperature. It could be  described by the dynamic form of the ideal gas law:

\begin{equation}\label{eq 14}  
\dot {V}_{b} = \frac{R_{i}}{M_{n}} \Biggl[ \dot {m}_{g} \frac{T_{g}}{P_{g}} + m_{g} \frac{\dot {T}_{g}}{P_{g}} \Biggr].
\end{equation}

The evolution of the ratio of rate in change of mass $\dot {m}_{g} \frac{T_{g}}{P_{g}}$ to change in temperature $m_{g} \frac{\dot {T}_{g}}{P_{g}}$, is plotted in Fig. \ref{model-resultsa}B. 
Throughout the vaporization process, the ratio $\biggl(\frac{\dot {m}_{g} \frac{T_{g}}{P_{g}}}{m_{g} \frac{\dot {T}_{g}}{P_{g}}}\biggr)$ remains significantly larger than one indicating that the heat is almost exclusively utilized in phase change with minimal contribution to gas heating.  

\subsection{Vaporization regimes}   \label{Dominant vaporization mechanisms}
\begin{figure*}[ht]
 \begin{center}
 \includegraphics[width=0.75\linewidth]{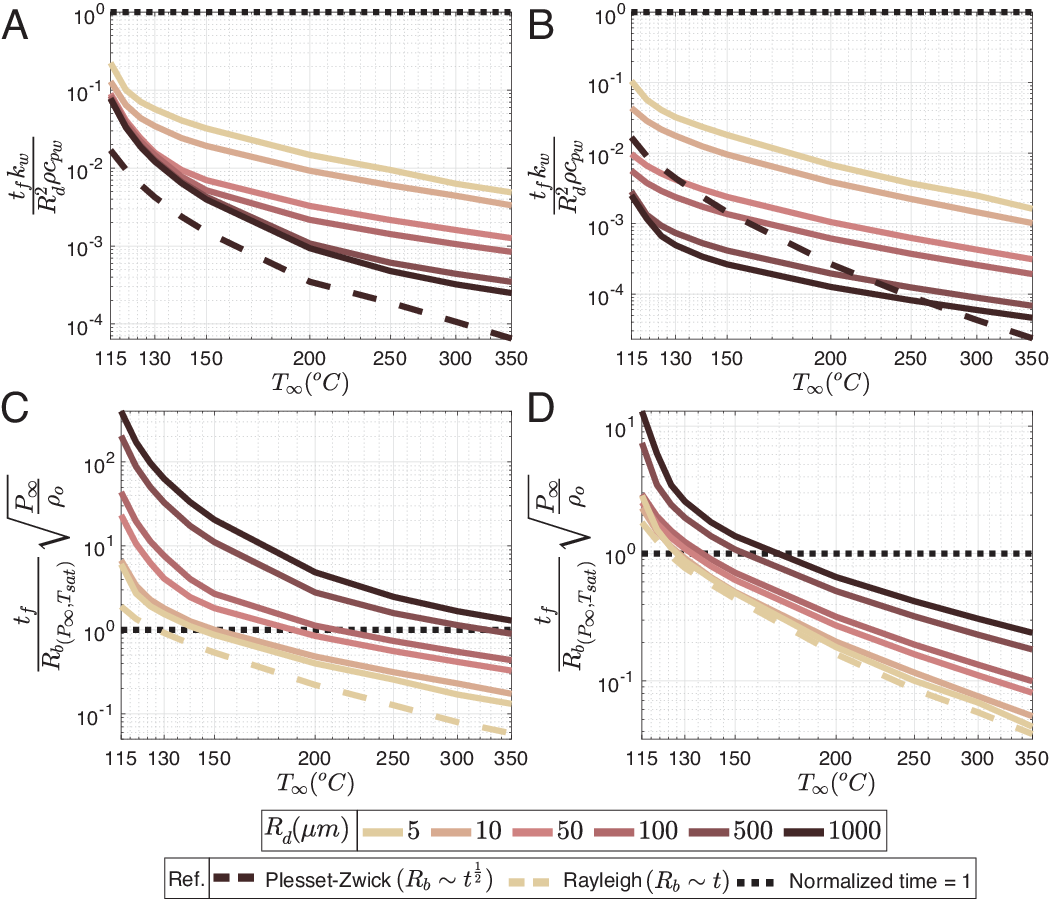} \\ 
 \caption{Dominant physical mechanism leading to bubble growth at the time of full vaporization $t_{f}$ as a function of the bath temperature. $t_{f}$ is normalized, in the top row, with the thermal-diffusion timescale, and in the bottom row, with the inertial time scale. A. and C. depict water, while the enthalpy of vaporization is reduced by a factor $\alpha $ $=$ $6.6$ in B. and D. to mimic pentane. Plesset-Zwick's and Rayleigh's models solved for equivalent bubble radius for $R_{d}$ of $1000$ $\mu m$ and $ 5$ $\mu m$ respectively, as a function of temperature.
 \label{dominant_mechanism}}
\end{center}
\end{figure*}

The dominant physical mechanisms for the final growth of the bubble phase depend on droplet size and ambient temperature (see section \ref{Model validation}). To characterize the specific physical phenomenon that governs the final stage of vaporization, we extract the time at which the droplet is fully vaporized $t_{f}$ and non-dimensionalize it with, on the one hand, the characteristic heat diffusion time scale$: \tau_{thermal} = \frac{R_{d}^{2} \rho_{o} c_{p_{w}}}{k_{w}}$ and, on the other hand, with the inertial time scale$: \tau_{inertial} = R_{b_{(P_{\infty}, T_{sat})}} \sqrt{\frac{\rho_{o}}{P{\infty}}}$ where $R_{b_{(P_{\infty}, T_{sat})}} = \sqrt[3]{\frac{\rho_{w} R_{d}^{3}}{\rho_{g_{(P_{\infty}, T_{sat})}}}}$. The former is shown in Fig. \ref{dominant_mechanism}A, and the latter in Fig. \ref{dominant_mechanism}C. For reference, we also plot analytical models: the Plesset-Zwick model \cite{plesset1954growth} describing thermal diffusive growth, and the Rayleigh model \cite{rayleigh1917pressure} that addresses the dominance of inertial effects.

As clarified in the previous section, these timescales are only rough estimates that help in understanding the main features. The absence of flat curves in this context results from the significant influence of ambient temperature influencing both diffusive and inertial timescales. Nonetheless, we can consider two limiting cases, the vaporization of large droplets (here $1$ $mm$) follows the Plesset-Zwick model, and is thus dominated by diffusion for the temperature range ($115 - 250$ $^{\circ}C$). Second, small droplets (here $5$ and $10$ $\mu m$) are entirely dominated by inertia and thus follow the Rayleigh model i.e. for low and medium degrees of superheat. Using these two curves as references, it is immediately evident that droplets under study encompass some droplet sizes whose vaporization dynamics are dominated neither by diffusion nor by inertia. 

This observation is further supported by the trends depicted in Fig. \ref{dominant_mechanism}B and Fig. \ref{dominant_mechanism}D where the vaporization enthalpy is reduced to that of Pentane, which corresponds to a factor $\alpha = 6.6$ (measured in $J.kg^{-1}$). The other liquid properties are kept unchanged. Inertia becomes increasingly dominant as the vaporization enthalpy decreases and extends further to cover higher temperatures while the thermal diffusive limit is pushed back to the low degree of superheat. %Substantial data that does not fall within inertial and thermal regimes may either be in transition from inertia to thermal or be dominated by surface tension. Data around the thermal diffusion limit are likely to be within the transitional region, whereas those above the inertial limit are likely to be governed predominantly by surface tension.

\subsection{Influence of the thermal diffusivity of the host liquid} \label{diffusivity}
Thermal diffusivity influences the bubble dynamics, as vaporization relies on heat transfer from the surrounding liquid. Avedisian and Suresh \cite{avedisian1985analysis} had proposed that the entire vaporization for a $1000$ $\mu m$ event splits into two stages, first during the initial bubble growth when the boundary layer is inside the droplet. In this way, thermal diffusivity and other properties of the outer liquid are of lesser importance. In the second stage, the thermal boundary layer develops outside of the droplet in the heating fluid, and thus its diffusivity has significance in driving the bubble growth.

 \begin{figure*}[ht]
 \begin{center}
 \includegraphics[width=0.8\linewidth]{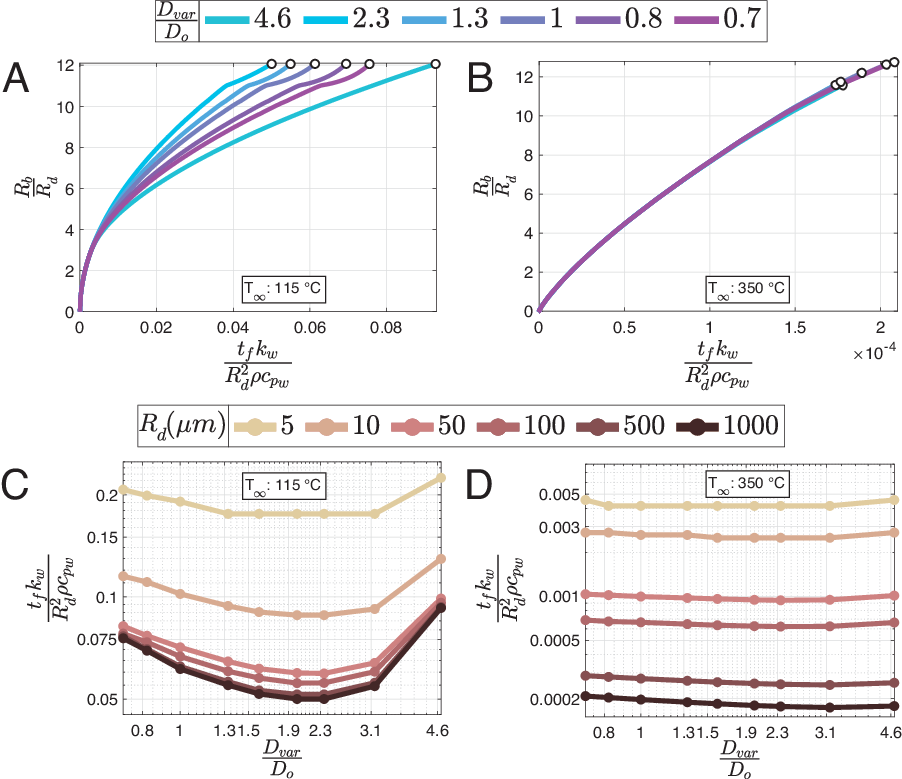} \\ 
 \caption{Effect of thermal diffusivity of the outer medium. Bubble growth profile of a $1000$ $\mu m$ radius droplet at temperatures A. $115$ $^{\circ}C$, B. $350$ $^{\circ}C$. Influence on full vaporization time across all ranges of sizes at C. $115$ $^{\circ}C$, D. $350$ $^{\circ}C$.
 \label{radius_diffusivity}}
\end{center}
\end{figure*}

We have plotted the bubble growth profile until full vaporization in Fig. \ref{radius_diffusivity} (A, B) at $115$ and $350$ $^\circ C$ respectively, for a $1000$ $\mu m$ radius droplet. These two temperatures establish the upper and lower limits of our simulations, effectively covering the entire range of dynamics. We focus on the time scale of thermal diffusion since its influence is under study here. Each figure shows the effect of varying oil diffusivity $(D_{var})$ non-dimensionalized by the original oil diffusivity $(D_{o})$. At $115$ $^\circ C$, two stages of growth are indeed visible in the plot: initially, all curves collapse for different diffusivity indicating no thermal influence of the outer medium. After a certain time, the curves start to diverge, highlighting the second growth stage and the dependence on the thermal properties of the outer medium. However, this is significant only for temperatures and droplet size ranges where vaporization is dominated by thermal diffusive growth: at the upper end of our temperature range ($350$ $^\circ C$) in Fig. \ref{radius_diffusivity}B complete vaporization occurs within the first stage without relying on heat diffusion from outside. All growth curves collapse till complete vaporization indicating no influence of diffusion from the outer medium and validating results shown in Fig. \ref{dominant_mechanism}A for the same size and temperature. 

An interesting observation is that the time to reach full vaporization shows a minimum as a function of the relative diffusivity. This effect is best seen in Fig. \ref{radius_diffusivity} (C and D) showing the time of full vaporization nondimensionalized by the characteristic time of heat diffusion. The location of the minimum is $\frac{D_{var}}{D_{o}}$ $\approx$ $2$ for low degrees of superheat while at high degrees of superheat, it is almost negligible. This indicates that at a $350$ $^\circ C$, there is enough energy stored in all size ranges of the droplet that it would vaporize fully without the necessity of acquiring heat from the outer medium. One could determine the temperature required ($T^{\prime}$) for complete droplet vaporization without depending on heat diffusion from the surrounding medium. This can be achieved through the equation $T^{\prime} = \frac{H_{w(P_{\infty}, T_{\infty})}}{C_{pw(P_{\infty}, T_{sat})}} + T_{sat}$, yielding values of $308$ $^o C$ where $H_{w}$ is calculated at $350$ $^o C$.
                    
\subsection{Influence of phase-change enthalpy} \label{Phase-change}

As seen in section \ref{Dominant vaporization mechanisms}, reducing phase change enthalpy shifts the vaporization regimes. We now explore its influence on the bubble's growth. We have scaled down the water enthalpy ($H_{w}$) by a factor $\alpha =$ $3.3$, $6.6$, or $13.2$. The value (measured in $J/kg$) obtained after reducing it by a factor of $6.6$ corresponds to pentane, whereas the value $14$ corresponds to bromoform. Most common liquids have a vaporization enthalpy ranging from $50 \%$ (e.g., pentane) to $100 \%$ of that of water. There are exceptional liquids with large molecules that have a lower enthalpy (as in the case of bromoform). 

\begin{figure*}[ht]
 \begin{center}
 \includegraphics[width=0.78\linewidth]{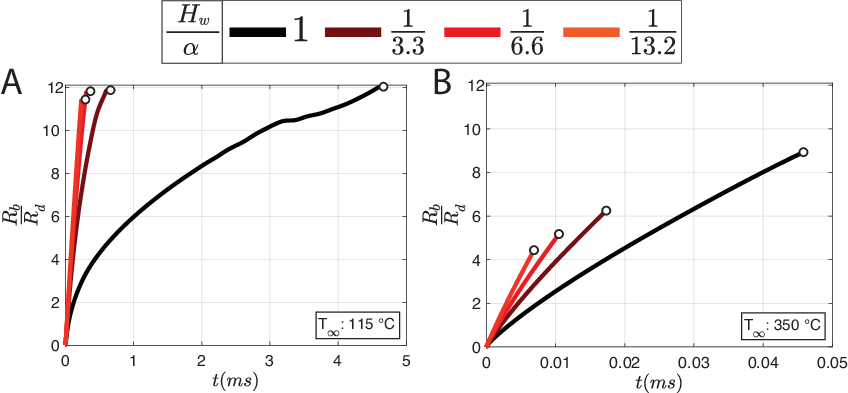} \\ 
 \caption{Effect of vaporization enthalpy: A - B. bubble growth at $115$ and $350$ $^{\circ}C$.
 \label{Diffusivity_Enthalpy}}
\end{center}
\end{figure*}

Fig. \ref{Diffusivity_Enthalpy} shows the influence of reducing the phase-change enthalpy. In (A, B) we have plotted the vapor bubble radius as a function of time at a bath temperature of $115$, and, $350$ $^\circ C$ respectively, for a $100$ $\mu m$ droplet. As phase-change enthalpy reduces, the conversion ratio from transferred heat to potential energy (i.e., vaporized mass and pressure) increases. As a result, vaporization happens much faster, and the effect is even more pronounced for low degrees of superheat. As observed in section \ref{Phase-change}, when the temperature is low the system contains relatively less stored heat, causing it to depend on heat diffusion from the surrounding environment. A reduction in phase change enthalpy necessitates a smaller amount of heat, resulting in faster vaporization and bubble growth.

The influence of phase change enthalpy is not limited to vaporization and affects the post-vaporization dynamics. With lower enthalpy at a high degree of superheat (Fig. \ref{Diffusivity_Enthalpy}B), vaporization occurs at a smaller bubble radius. The excess energy is then utilized to raise gas pressure, leading to fast growth dynamics since no further energy is needed for phase change. This rapid expansion amplifies inertial effects, facing substantial resistance from the surrounding medium.

\subsection{Post-vaporization damped oscillations} \label{post-vap}

 \begin{figure*}[ht]
 \begin{center}
B \includegraphics[width=0.7\linewidth]{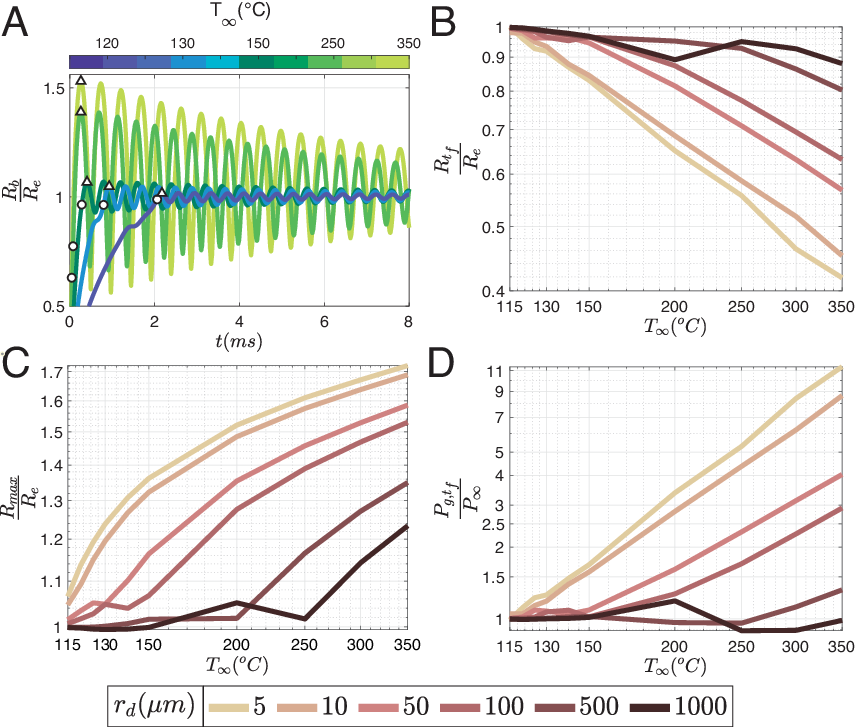} \\ 
 \caption{Post-vaporization dynamics. A. Time evolution of bubble oscillations non-dimensionalized by the equilibrium radius for a $100$ $\mu m$ radius droplet at a range of ambient temperatures. The white circles indicate the radius at full vaporization ($R_{t_{f}}$), while the white triangle indicates the maximum radius ($R_{max}$). For a range of ambient temperatures and droplet sizes: B. radius at the time of full vaporization non-dimensionalized by the equilibrium radius, C. maximum bubble radius non-dimensionalized by the equilibrium radius, D. gas pressure at the time of full vaporization non-dimensionalized by the ambient pressure.
 \label{Postvaporization}}
\end{center}
\end{figure*}

Once the droplet is fully vaporized, the resulting bubble overshoots its equilibrium radius $R_{e}^{3} = \frac{3 \rho_{w} V_{d_{0}} R_{i}T_{\infty}}{4 \pi P_{\infty} M_{w}}$ as can be seen in Fig. \ref{Postvaporization}A, where the bubble radius is non-dimensionalized by its equilibrium radius, and the vaporization dynamics is studied for ambient temperatures ranging from 120 $^o$C to 350 $^o$C. The overshoot is driven by the difference between gas pressure and ambient pressure, resulting from the potential energy available at the time of complete vaporization $t_{f}$. A decrease in temperature leads to a reduction in energy, thereby resulting in a decrease in the overshoot radius $R_{max}$.

At a high degree of superheat vaporization is accomplished at a radius $R_{t_{f}}$ smaller than the equilibrium radius, due to the excess amount of energy in the system. For instance, vaporization of $10$ $ \mu m$ droplet at $350$ $^{\circ}C$ occurs at a radius $0.45$ times smaller than its equilibrium radius as shown in Fig. \ref{Postvaporization}B, where the radius at the time of complete vaporization is plotted for different droplet sizes. Here vaporization completes at a temperature well above the saturation temperature: for $10$ $ \mu m$  at $350$ $^{\circ}C$ the residual pressure at complete vaporization is $8.6$ times higher than the ambient pressure, thus driving the bubble after vaporization towards an overshoot of $1.68$ times the equilibrium radius, see Fig. \ref{Postvaporization} (C-D). On the other hand, for small-size droplets at a low degree of superheat, the bubble radius at full vaporization is nearly equal to the equilibrium radius with no energy to drive an overshoot and is therefore negligible. 

The initial bubble expansion follows well-known damped oscillation behavior. In the model, damping has been described in terms of acoustic reradiation, and viscous components. Assuming small radial excursion one may find the eigenfrequency $(\omega_{o})$ as well as damping $(\zeta_{tot})$ from the linearized equation of motion given as \cite{versluis2020ultrasound}:

\begin{equation}\label{eq 35}
\omega_{o} = \sqrt{ \frac{3 \gamma P_{\infty}}{\rho_{o} R_{e}^2} + (3\gamma - 1) \frac{2 \sigma_{o} }{\rho_{o} R_{e}^3}}, 
\end{equation}

\begin{equation}\label{eq 36}
\zeta_{tot} = \frac{3 \gamma P_{\infty}}{2 \rho_{o} c_{o} R_{e} \omega_{o}} + \frac{3\gamma \sigma_{o} }{\rho_{o} c_{o} R_{e}^2 \omega_{o}} + \frac{2\mu_{o}}{\rho_{o} R_{e}^2 \omega_{o}},
\end{equation}

where $\gamma$ indicates the polytropic exponent corresponding to 1 for our isothermal system and $\sigma_{o}$ is the surface tension of oil. The frequency estimated from the Fourier transform of the simulated curves falls very close to resonance frequency $f_{res} = f_{\circ} \sqrt{1-\zeta_{tot}^{2}} $ estimated from the eigenfrequency in Eq. \ref{eq 35}, with a damping contribution from Eq. \ref{eq 36}. The frequency remains close to the well-established Minnaert frequency with a $1/R_{b}$ dependency. In contrast, the frequency of a vapor bubble oscillating in an infinite liquid scales as $\sim 1/R_{b}^{2/3}$ \cite{prosperetti2017vapor} as a consequence of gas dissolution and/or vapor condensation. In our system, we assume a pure water vapor bubble oscillating in oil, with no gas available to dissolve. Additionally, at the highest re-compression level, condensation remains absent as it necessitates releasing a significant amount of energy into a medium that is already superheated. The effect of ambient temperature on frequency can be seen by a shift of a few kiloHertz with an increase in temperature, while the shift becomes smaller with an increase in droplet radius. Equilibrium bubble sizes in our system range from $64$ $\mu m$ to $12$ $mm$ depending on droplet size and ambient temperature. These bubble sizes, while driven acoustically, are mainly dominated by sound re-radiation \cite{prosperetti1977thermal} and thus have a large contribution in Eq. \ref{eq 36} as compared to the viscous effect (third term). These findings also apply to the current system, wherein no external force is applied, and oscillations result only from a vaporization impulse (see SI Fig. \ref{PV}).

\section{Discussion on models comparison in literature}
\label{sec:sample5}
%The present study focuses on post-nucleation dynamics, neglecting consideration of the energy needed for nucleation due to its minimal influence on bubble growth, as the energy is recovered once the nucleation threshold is surpassed \cite{karthika2016review}. We have used the convection-diffusion equation to better resolve both temporal and spatial interdependencies, improving the ability to capture heat transfer dynamics. Our findings closely correspond with estimations derived from the thickness of the thermal boundary layer, indicating either a thick boundary layer in the absence of external heat diffusion or a thin boundary layer where vaporization relies on external heat. However, this approach is expensive in terms of computational costs, and requires a large number of grid points, offering limited improvement in accuracy. Moreover, making the model dimensionless would not only enhance its robustness, but also reduce the computational cost.

In literature, the closest models representing the system are those proposed by Avedisian and Suresh \cite{avedisian1985analysis}, and Emery et al. \cite{emery2018bubble} that focus on droplets with a radius ranging from sub-millimeter to millimeter, and Roesle and Kulacki \cite{roesle2010boiling} focus on smaller droplets (with a range of $2$ - $15 \mu m$). The numerical solutions employed in these studies are also different: the first two utilize Landau immobilization, fixing the positions of moving interfaces, whereas the third implicitly solves the convection-diffusion equation to determine the temperature profile. 
%Emery et al.'s model has limited comparability as it neglects elementary physical effects such as surface energy, kinetic energy, inertia, viscosity, and acoustic effects during bubble growth. 
Emery et al.'s model has limited comparability as it neglects momentum conservation. As such, it cannot capture initial rapid vaporization dynamics. Our model's findings are consistent with the predictions of Avedisian and Suresh \cite{avedisian1985analysis} and Roesle and Kulacki \cite{roesle2010boiling} regarding the influence of the outer medium on vaporization, as discussed in Section \ref{diffusivity}. While these studies maintain the gas temperature at the boiling temperature throughout the simulations, the bubble growth dynamics presented here are more reliable because our model incorporates the time-dependent convective effects, gas pressure, as well as heat and mass transfer. Furthermore, our model predicts post-vaporization damped oscillations, as we also consider the acoustic damping term.

\section{Conclusion}
\label{sec:sample6}
In this article, we have modeled and numerically solved the vaporization dynamics of a super-heated water droplet surrounded by an infinite oil (octadecene) medium. The droplet size as well as the degree of supereat has been varied to understand their effect on vaporization and bubble growth dynamics. For our system, it was found that the conventional two-phase analytical models are only valid for millimeter-sized droplets vaporizing at a low degree of superheat. With a decrease in droplet size and an increase in superheat, the physical mechanisms dominating bubble growth are inertia and thermal diffusion. A crucial thermal property that determines the dominant mechanism is the phase-change enthalpy. Thermal diffusivity is only important at low temperatures where vaporization relies on heat diffusion from the surrounding liquids at high temperatures as the droplet has stored enough thermal energy. Finally, bubble oscillations following complete vaporization depend on the achieved gas pressure and is influenced by the ambient temperature of the oil bath. The resonance frequency of these oscillations is close to the Minnaert frequency of a free gas bubble in an infinite medium with a $(1/R_{b})$  dependency on size, unlike the oscillations of a pure vapor bubble where the resonance frequency scales with $1/R_{b}^{2/3}$.

\section*{Acknowledgments}
This work is funded by the Dutch Research Council grant (Veni AES 2018 - 16879). G. Lajoinie acknowledges funding from the European Research Council (ERC-2022-STG Super-692 FALCON, No. 101076844).

\begin{appendix}

\section{Theory: Model of the bubble growth}
\label{sec:sample:appendix1}
\subsection{Bubble dynamics}
The Navier–Stokes equations express the conservation of mass and momentum:

\begin{equation}\label{N-S eq}
\rho(\frac{\partial \boldsymbol{v}}{\partial t} + (\boldsymbol{v} . \nabla)\boldsymbol{v}) + \nabla P= \mu  \Delta \boldsymbol{v},
\end{equation}

\begin{equation}\label{Continuity}
\frac{\partial \rho}{\partial t} + \nabla . (\rho \boldsymbol{v}) = 0,
\end{equation}

Considering the flow is symmetric $(\boldsymbol{v} = v(r)\boldsymbol{e_{r}})$ and incompressible $(\frac{\partial \rho}{\partial t} = 0)$, with potential flow approximation $(\boldsymbol{v} = \nabla\phi) )$ the above equation becomes:

\begin{equation}\label{eq 300}
\rho(\nabla (\dot{\phi}) + \frac{1}{2} \nabla  (\nabla (\phi) . \nabla (\phi)) + \nabla P= \mu  \nabla (\Delta \phi ).
\end{equation}

Integrating equation $\ref{eq 300}$ in both drop and accompanying outer medium: 

\begin{equation} \label{eq 120}  %eq 20
 \begin{cases*}
   \rho_{2}\left((\dot{\phi}_{R_{2}} -\dot{\phi}_{R_{1}}) + \frac{1}{2} (\dot{R}_{2}^2 -\dot{R}_{1}^2)\right) + P_{R_{2}^-} -P_{R_{1}^+} = 0  & drop, - (a)\\
  \rho_{2}\left(-\dot{\phi}_{R_{2}} - \frac{1}{2} \dot{R}_2^2\right) + P_{\infty}-P_{R_{2}^+} = 0,  & medium. - (b)
\end{cases*}
\end{equation}

With spherical symmetry, the Laplacian simplifies to:

\begin{equation}\label{eq 400}
\Delta \phi = \frac{1}{r^2} \frac{\partial}{\partial r}(r^2 \frac{\partial \phi}{\partial r})   = 0,
\end{equation}

Substituting $\Delta\phi=\phi^{\prime}$, and $\frac{\partial}{\partial r}(r^2 \frac{\partial \phi}{\partial r})=\alpha$: 
 
\begin{equation}\label{eq 500}
\phi^{\prime}  = \frac{\alpha}{r^2}, 
\end{equation}

\begin{equation}\label{eq 600}
\phi = -\frac{\alpha}{r} + \beta, 
\end{equation}

In oil, the bubble has no long-range effect, so as  $\phi \rightarrow 0$, $r \rightarrow \infty$, hence $\beta = 0$. As $\dot{R} = \phi^{\prime}(R)$, $\alpha = \dot{R} R^2$. Solving incomprehensibility in both media and matching velocity at the interface:

\begin{equation}\label{eq 700}
\phi(r,t)  = -\frac {\dot{R_{1}} R_{1}^2}{r} = -\frac {\dot{R_{2}} R_{2}^2}{r}, 
\end{equation}

\begin{equation}\label{eq 800}
\dot{\phi}(r)  = -\frac{\ddot{R}_{2}R_{2}^2+2\dot{R}_{2}^2R_{2}}{r} = -\frac{\ddot{R}_{1}R_{1}^2+2\dot{R}_{1}^2R_{1}}{r},
\end{equation}

\begin{equation}\label{eq 1000}
\dot{\phi} (R_{1}) = -\frac{R_{1}}{R_{1}}(\ddot{R}_{1}R_{1}+2\dot{R}_{1}^2) = -\ddot{R}_{1}R_{1}-2\dot{R}_{1}^2, 
\end{equation}

\begin{equation}\label{eq 900}
\dot{\phi} (R_{2}) = -\frac{\ddot{R}_{2}R_{2}^2+2\dot{R}_{2}^2R_{2}}{R_{2}} = -\ddot{R}_{2}R_{2}-2\dot{R}_{2}^2 = -\frac{R_{1}}{R_{2}}(\ddot{R}_{1}R_{1}+2\dot{R}_{1}^2).
\end{equation}

The normal stress balance at second interfaces could be written as:
\begin{equation}\label{eq 130}
2\mu_{1} \frac{\partial v}{\partial r} \bigg\rvert_{R} - 2\mu_{2} \frac{\partial v}{\partial r} \bigg\rvert_{R}  + P_{R_{2}^+} - P_{R_{2}^-} = -\delta P_{laplace},
\end{equation}

Where:
\begin{equation}\label{eq 140}
\delta P_{laplace} = \kappa \sigma = \frac{2 \sigma_{12}}{R}, 
\end{equation}

\begin{equation}\label{eq 150}
v = \frac{\partial \phi}{\partial r} = - \frac{\partial}{\partial r}  \left(\frac{\dot{R} R^2}{r}\right) = \frac{\dot{R} R^2}{r^2},
\end{equation}

\begin{equation}\label{eq 160}
\frac{\partial v}{\partial r} \bigg\rvert_{R} = \frac{\partial \phi}{\partial r}  \left(\frac{\dot{R} R^2}{r^2}\right) \bigg\rvert_{R} = \frac{-2 \dot{R}}{R},
\end{equation}

\begin{equation}\label{eq 170}
2\mu_{1} \frac{\partial v}{\partial r} \bigg\rvert_{R} - 2\mu_{2} \frac{\partial v}{\partial r} \bigg\rvert_{R} = 2 \frac{\partial v}{\partial r} \bigg\rvert_{R} (\mu_{1}-\mu_{2}) =  \frac{\dot{R}}{R}  4 (\mu_{2}-\mu_{1}). 
\end{equation}

Rewriting the normal stress balances at two interfaces:

\begin{equation} \label{eq 180}  %eq 21
 \begin{cases*}
   P_{R_{1}^+} = P_{g} - \frac{2 \sigma_{1}}{R_{1}}  - \frac{\dot{R_{1}}}{R_{1}} 4 \mu_{1},  & first, - (a)\\
P_{R_{2}^+} = P_{R_{2}^-} - \frac{2 \sigma_{12}}{R_{2}}  + \frac{\dot{R_{2}}}{R_{2}} 4 (\mu_{1}-\mu_{2}) , &  second. - (b)
\end{cases*}
\end{equation}

The expression for a spherical bubble expanding within a droplet, denoted as equation $\ref{eq 190} (a)$, is derived by substituting equations $\ref{eq 900}$, $\ref{eq 1000}$, and $\ref{eq 180} (a)$ into equation $\ref{eq 120} (a)$. Similarly, the expression for the growth of the bubble from the droplet to infinity, represented as equation $\ref{eq 190} (b)$, is obtained by substituting equations $\ref{eq 900}$ and $\ref{eq 180} (b)$ into equation $\ref{eq 120} (b)$.

\begin{equation} \label{eq 190}  %eq 22
 \begin{cases*}
    \rho_{1}\left((1-\frac{R_{1}}{R_{2}})(\ddot{R}_{1}R_{1}+2\dot{R}_{1}^2) + \frac{\dot{R}_{2}^2 -\dot{R}_{1}^2}{2} \right) 
 = P_{g} - \frac{2 \sigma_{1}}{R_{1}}  - \frac{\dot{R_{1}}}{R_{1}} 4 \mu_{1} - P_{R_{2}^-},  & (a)\\
\rho_{2}\left(\frac{R_{1}}{R_{2}}(\ddot{R}_{1}R_{1}+2\dot{R}_{1}^2) - \frac{\dot{{R}_2}^2}{2} \right)  = P_{R_{2}^-} - \frac{2 \sigma_{12}}{R_{2}}  + \frac{\dot{R_{2}}}{R_{2}} 4 (\mu_{1}-\mu_{2})   - P_{\infty}. &  (b)
\end{cases*}
\end{equation}

Finally, the equation describing the dynamics of the bubble in two mediums is obtained by adding eq \ref{eq 190} $(a)$ and $(b)$:

\begin{equation} \label{eq 230}
\begin{split}
(\ddot{R}_{1}R_{1}+2\dot{R}_{1}^2) \left(\rho_{2}-\rho_{1}) \frac{R_{1}}{R_{2}} + \rho_{1}\right) + \frac{(\rho_{1}-\rho_{2}) \dot{R}_{2}^2 }{2} - \frac{\rho_{1} \dot{R}_{1}^2}{2} \\ = P_{g} + \frac{R_{1}}{c_{o}} \dot{P_{g}} - P_{\infty} -2\left(\frac{ \sigma_{12}}{R_{2}} + \frac{ \sigma_{1}}{R_{1}}\right) -4\left( \frac{\dot{R_{2}}}{R_{2}} (\mu_{2}-\mu_{1}) +\frac{\dot{R_{1}}}{R_{1}} \mu_{1} \right). 
\end{split}
\end{equation}

\subsection{Vaporization dynamics}
The pressure driving the bubble in accordance with temperature is given by semi-empirical Antoine law which relates as:

\begin{equation}\label{eq 240}
P_{g}  = 10^ {5+A - \frac{B}{C+T_{g}}},
\end{equation}
where, $A, B, and C$ are the Antoine coefficients and $T_{g}$ represents temperature of the gas. In dynamic form:

\begin{equation}\label{eq 2500}
\frac{\dot{P_{g}}}{P_{g}} = \frac{\dot{T_{g}}Bln(10)}{(C+T_{g})^2}.
\end{equation}

During vaporization, the bubble surface and the neighboring liquid motion are influenced by the mass transfer from liquid to gas. The velocity $\dot{R_{b}}$ on the bubble side of the interface is higher than the velocity on the liquid side $\dot{R_{1}}$ by an amount that could be estimated by the rate at which mass flows per unit volume.

\begin{equation}\label{eq 2600}
\dot{R_{1}} = \dot{R_{b}} - \frac{\dot{m}}{4\pi R_{1}^2 \rho_{1}},
\end{equation}

\begin{equation}\label{eq 2700}
\frac{\dot{R_{1}}}{R_{1}} =  \frac{\dot{R_{b}}}{R_{1}} - \frac{\dot{m}}{3m} \frac{\rho_{g}}{\rho_{1}},
\end{equation}

The heat provided to the system is utilized in raising the gas temperature and providing phase change enthalpy:

\begin{equation}\label{eq 2800}
{4\pi K_{1} R_{1}^2} \frac{\partial T}{\partial r} \bigg|_{R_{1}} = mC_{pg} \dot{T_{g}} + \dot{m}H_{v},
\end{equation}

\begin{equation}\label{eq 2800a}
\dot{m} = \frac{{4\pi K_{1} R_{1}^2}}{H_{v}}  \frac{\partial T}{\partial r} \bigg|_{R_{1}} - \frac{mC_{pg} \dot{T_{g}}}{H_{v}},   
\end{equation}

\begin{equation}\label{eq 2900}
\frac{\dot{m}}{m} = \frac{{4\pi K_{1} R_{1}^2}}{H_{v}m}  \frac{\partial T}{\partial r} \bigg|_{R_{1}} - \frac{C_{pg} \dot{T_{g}}}{H_{v}}.   
\end{equation}

A better estimate of gas temperature can be achieved by employing the dynamic form of the ideal gas law as the state function. This approach takes into account the variations in radius, pressure, mass, and temperature over time, providing a comprehensive understanding of how these factors evolve dynamically. 

\begin{equation}\label{eq 3000}
\frac{3 \dot{R_{b}}}{R_{b}} + \frac{ \dot{P_{g}}}{P_{g}} = \frac{ \dot{m}}{m} + \frac{ \dot{T_{g}}}{T_{g}}, 
\end{equation}

Where $R_{b} = R_{1}$. The equation that governs the dynamic gas temperature is finally obtained by substituting equation \ref{eq 2500}, \ref{eq 2700}, and \ref{eq 2900}  in \ref{eq 3000}:  

\begin{equation}\label{eq 3100}
\frac{ \dot{T_{g}}}{T_{g}} - \frac{ \dot{P_{g}}}{P_{g}} = \frac{3 \dot{R_{1}}}{R_{1}} - \frac{\dot{m}}{m}\Big(1+\frac{\rho_{g}}{\rho_{1}}\Big),
\end{equation}

\begin{equation}\label{eq 3200}
\dot{T_{g}} \Bigg[\frac{1}{T_{g}} - \frac{Bln(10)}{(C+T_{g})^2} - \frac{C_{pg} }{H_{v}}\Big(1+\frac{\rho_{g}}{\rho_{1}}\Big) \Bigg] =  \frac{3 \dot{R_{b}}}{R_{1}} - \frac{{4\pi K_{1} R_{1}^2}}{H_{v}m}  \frac{\partial T}{\partial r} \bigg|_{R_{1}} \Big(1+\frac{\rho_{g}}{\rho_{1}}\Big).
\end{equation}

Once the water is fully vaporized the transfer of mass $\dot{m}$ from the liquid phase to the vapor phase becomes zero and the heat is utilized only to raise the gas temperature. In this stage, the gas pressure is calculated by equation \ref{eq 3000}, and its temperature is calculated by

\begin{equation}\label{eq 3300}
\dot{T_{g}}= \frac{{4\pi K_{1} R_{1}^2}}{C_{pg}m}  \frac{\partial T}{\partial r} \bigg|_{R_{1}}.
\end{equation}

\end{appendix} 

\section*{Supplementary information}
\subsection*{Post-vaporization damped oscillations}

  \begin{figure*}[ht]
 \begin{center}
 \includegraphics[width=1\linewidth]{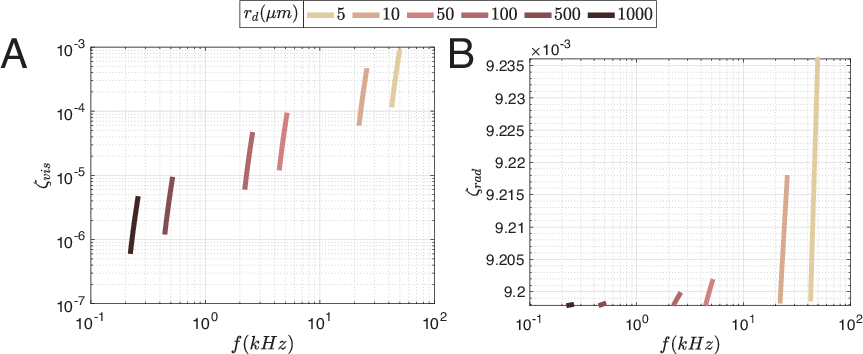} \\ 
 \caption{Damping coefficient of bubble oscillation as a function of frequency across different droplet sizes. A. Viscous damping, B. Acoustic damping. Increased damping for each size corresponds to the rise in ambient temperature from 115 to 350$^{o}$ C. 
 \label{PV}}
\end{center}
\end{figure*}

\newpage
 \bibliographystyle{elsarticle-num} 
 \bibliography{elsarticle-template-num}

\begin{thebibliography}{10}
\expandafter\ifx\csname url\endcsname\relax
  \def\url#1{\texttt{#1}}\fi
\expandafter\ifx\csname urlprefix\endcsname\relax\def\urlprefix{URL }\fi
\expandafter\ifx\csname href\endcsname\relax
  \def\href#1#2{#2} \def\path#1{#1}\fi

\bibitem{letan1968mechanism}
R.~Letan, E.~Kehat, The mechanism of heat transfer in a spray column heat exchanger, AIChE Journal 14~(3) (1968) 398--405.

\bibitem{mccabe1993unit}
W.~L. McCabe, J.~C. Smith, P.~Harriott, Unit operations of chemical engineering, Vol.~5, McGraw-hill New York, 1993.

\bibitem{can2012level}
E.~Can, A.~Prosperetti, A level set method for vapor bubble dynamics, Journal of Computational Physics 231~(4) (2012) 1533--1552.

\bibitem{ory2000growth}
E.~Ory, H.~Yuan, A.~Prosperetti, S.~Popinet, S.~Zaleski, Growth and collapse of a vapor bubble in a narrow tube, Physics of Fluids 12~(6) (2000) 1268--1277.

\bibitem{lee1996spherical}
H.~S. Lee, H.~Merte~Jr, Spherical vapor bubble growth in uniformly superheated liquids, International Journal of Heat and Mass Transfer 39~(12) (1996) 2427--2447.

\bibitem{dergarabedian1953rate}
P.~Dergarabedian, The rate of growth of vapor bubbles in superheated water, Journal of Applied Mechanics 20~(4) (2021) 537--545.

\bibitem{kosky1968bubble}
P.~Kosky, Bubble growth measurements in uniformly superheated liquids, Chemical Engineering Science 23~(7) (1968) 695--706.

\bibitem{florschuetz1969growth}
L.~Florschuetz, C.~Henry, A.~R. Khan, Growth rates of free vapor bubbles in liquids at uniform superheats under normal and zero gravity conditions, International Journal of Heat and Mass Transfer 12~(11) (1969) 1465--1489.

\bibitem{rayleigh1917pressure}
L.~Rayleigh, On the pressure developed in a liquid during the collapse of a spherical cavity, Philosophical Magazine 6~(34) (1917) 94--98.

\bibitem{plesset1954growth}
M.~S. Plesset, S.~A. Zwick, The growth of vapor bubbles in superheated liquids, Journal of Applied Physics 25~(4) (1954) 493--500.

\bibitem{chernov2020new}
A.~Chernov, A.~Pil’nik, I.~Vladyko, S.~Lezhnin, New semi-analytical solution of the problem of vapor bubble growth in superheated liquid, Scientific Reports 10~(1) (2020) 16526.

\bibitem{sideman1964direct}
S.~Sideman, Y.~Taitel, Direct-contact heat transfer with change of phase: evaporation of drops in an immiscible liquid medium, International Journal of Heat and Mass Transfer 7~(11) (1964) 1273--1289.

\bibitem{tochitani1977vaporization}
Y.~Tochitani, Y.~Mori, K.~Komotori, Vaporization of single liquid drops in an immiscible liquid part i: Forms and motions of vaporizing drops, W{\"a}rme-und Stoff{\"u}bertragung 10~(1) (1977) 51--59.

\bibitem{tochitani1977vaporizationa}
Y.~Tochitani, T.~Nakagawa, Y.~Mori, K.~Komotori, Vaporization of single liquid drops in an immiscible liquid part ii: Heat transfer characteristics, W{\"a}rme-und Stoff{\"u}bertragung 10~(2) (1977) 71--79.

\bibitem{lajoinie2014ultrafast}
G.~Lajoinie, E.~Gelderblom, C.~Chlon, M.~B{\"o}hmer, W.~Steenbergen, N.~De~Jong, S.~Manohar, M.~Versluis, Ultrafast vapourization dynamics of laser-activated polymeric microcapsules, Nature Communications 5~(1) (2014) 3671.

\bibitem{lajoinie2020three}
G.~Lajoinie, M.~Visscher, E.~Blazejewski, G.~Veldhuis, M.~Versluis, Three-phase vaporization theory for laser-activated microcapsules, Photoacoustics 19 (2020) 100185.

\bibitem{avedisian1985analysis}
C.~Avedisian, K.~Suresh, Analysis of non-explosive bubble growth within a superheated liquid droplet suspended in an immiscible liquid, Chemical Engineering Science 40~(12) (1985) 2249--2259.

\bibitem{roesle2010boiling}
M.~Roesle, F.~Kulacki, Boiling of small droplets, International Journal of Heat and Mass Transfer 53~(23-24) (2010) 5587--5595.

\bibitem{emery2018bubble}
T.~S. Emery, P.~A. Raghupathi, S.~G. Kandlikar, Bubble growth inside an evaporating liquid droplet introduced in an immiscible superheated liquid, International Journal of Heat and Mass Transfer 127 (2018) 313--321.

\bibitem{plesset1949dynamics}
M.~S. Plesset, The dynamics of cavitation bubbles, Journal of Applied Mechanics 16~(3) (2021) 277--282.

\bibitem{antoine1888nouvelle}
M.~C. Antoine, Nouvelle relation entre les tensions et les temperatures, Comptes Rendus des Séances de l'Académie des Sciences 107 (1888) 681--684.

\bibitem{nistlink}
https://webbook.nist.gov.

\bibitem{STocta}
https://materials.springer.com/lb/docs/sm\_lbs\_978-3-540-69409-0\_2.

\bibitem{mikic1970bubble}
B.~Mikic, W.~Rohsenow, P.~Griffith, On bubble growth rates, International Journal of Heat and Mass Transfer 13~(4) (1970) 657--666.

\bibitem{versluis2020ultrasound}
M.~Versluis, E.~Stride, G.~Lajoinie, B.~Dollet, T.~Segers, Ultrasound contrast agent modeling: a review, Ultrasound in Medicine \& Biology 46~(9) (2020) 2117--2144.

\bibitem{prosperetti2017vapor}
A.~Prosperetti, Vapor bubbles, Annual Review of Fluid Mechanics 49 (2017) 221--248.

\bibitem{prosperetti1977thermal}
A.~Prosperetti, Thermal effects and damping mechanisms in the forced radial oscillations of gas bubbles in liquids, The Journal of the Acoustical Society of America 61~(1) (1977) 17--27.

\end{thebibliography}





% \end{thebibliography}
\end{document}